\documentclass[axioms,review,accept,10pt,a4paper]{mdpi}

\firstpage{1} 
\articlenumber{x}
\doinum{10.3390/------}
\pubvolume{5}
\pubyear{2016}
\copyrightyear{2016}
\externaleditor{Academic Editor: {Hans J. Haubold}}
\history{Received: 8 July 2016 ; Accepted: 8 August 2016; Published: date}

\usepackage{soul, microtype}
\newcommand\myurl[1]{\changeurlcolor{black}\url{#1}\changeurlcolor{blue}}

\newcommand{\highlight}[1]{\colorbox{yellow}{#1}}

 \theoremstyle{mdpi}
 \newcounter{thm}
 \newcounter{ex}
 \newcounter{re}
 \theoremstyle{mdpidefinition}

\Title{Approach of Complexity in Nature: Entropic~Nonuniqueness}

\Author{\hl{Constantino Tsallis} $^{1,2}$}
\AuthorNames{Constantino Tsallis}

\address{%
$^{1}$ \quad Centro Brasileiro de Pesquisas Fisicas, and National Institute for Science and Technology for Complex Systems, Rua Xavier Sigaud 150, Rio de Janeiro-RJ 22290-180, Brazil; tsallis@cbpf.br; \hl{Tel.: +55-21-21417190}\\
$^{2}$ \quad Santa Fe Institute, 1399 Hyde Park Road, Santa Fe, NM 87501, USA}

\abstract{Boltzmann introduced in the 1870s a logarithmic measure for the connection between the thermodynamical entropy and the probabilities of the microscopic configurations of the system. His celebrated entropic functional for classical systems was then extended by Gibbs to the entire phase space of a many-body system and by von Neumann in order to cover quantum systems, as~well. Finally, it was used by Shannon within the theory of information. The simplest expression of this functional corresponds to a discrete set of $W$ microscopic possibilities and is given by $S_{BG}= -k\sum_{i=1}^W p_i \ln p_i$ ($k$ is a positive universal constant; {BG} stands for {Boltzmann--Gibbs}). This~relation enables the construction of BG
 statistical mechanics, which, together with the Maxwell equations and classical, quantum and relativistic mechanics, constitutes one of the pillars of contemporary physics. The BG theory has provided uncountable important applications in physics, chemistry, computational sciences, economics, biology, networks and others. As argued in the textbooks, its application in physical systems is legitimate whenever the hypothesis of {ergodicity} is satisfied, i.e., when ensemble and time averages coincide. However, {what can we do when ergodicity and similar simple hypotheses are violated}, which indeed happens in very many natural, artificial and social complex systems. The possibility of generalizing BG statistical mechanics through a family of non-additive entropies was advanced in 1988, namely $S_q=k\frac{1-\sum_{i=1}^W p_i^q}{q-1}$, which recovers the additive $S_{BG}$ entropy in the $q$~$\to$~1~limit. 
The index $q$ is to be determined from mechanical first principles, corresponding to complexity universality classes. Along three decades, this idea intensively evolved world-wide (see the Bibliography in {http://tsallis.cat.cbpf.br/biblio.htm}) and led to a plethora of predictions, verifications and applications in physical systems and elsewhere. As expected, whenever a {paradigm shift} is explored, some controversy naturally emerged, as well, in the community. The present status of the general picture is here described, starting from its dynamical and thermodynamical foundations and ending with its most recent physical applications.
}


\begin{document}

\section{Introduction}

In light of contemporary physics, the qualitative and quantitative study of nature may be done at various levels, which here we refer to as {microcosmos}, {mesocosmos} and {macrocosmos}. At the macroscopic level, we have thermodynamics; at the microscopic level, we have mechanics (classical, quantum, relativistic mechanics, quantum chromodynamics) and the laws of electromagnetism, which~enable in principle the full description of all of the degrees of freedom of the system; at the mesoscopic level, we~focus on the degrees of freedom of a typical particle, representing, in one way or another, the~behavior of most of the degrees of freedom of the system. The laws that govern the microcosmos together with theory of probabilities are the basic constituents of statistical mechanics, a theory, which then establishes the connections between these three levels of description of nature. At the microscopic level, we typically address classical or quantum equations of evolution with time, trajectories in phase space, Hamiltonians, Lagrangians, among other mathematical objects. At the mesoscopic level, we address Langevin-like, master-like and Fokker--Planck-like equations. Finally, at the macroscopic level, we address the laws of thermodynamics with its concomitant Legendre transformations between the appropriate variables. 

In all of these theoretical approaches, the thermodynamical entropy $S$, introduced by Clausius in 1865 \cite{Clausius1865} and its corresponding entropic functional $S(\{ p_i \})$ play a central role. In a stroke of genius, the first adequate entropic functional was introduced (for what we nowadays call classical systems) by Boltzmann in the 1870s \cite{Boltzmann1872,Boltzmann1877} for a one-body phase space and was later on extended by Gibbs \cite{Gibbs1902} to the entire many-body phase space. Half a century later, in 1932, von Neumann \cite{vonNeumann1932} extended the Boltzmann--Gibbs (BG) entropic functional to quantum systems. Finally, in 1942, Shannon showed \cite{Shannon1948} the crucial role that this functional plays in the theory of communication. The simplest expression of this functional is that corresponding to a single discrete random variable admitting $W$ possibilities with nonvanishing probabilities $\{p_i\}$, namely:
\begin{equation}
S_{BG}=-k \sum_{i=1}^W p_i \ln p_i \;\;\; \Bigl(\sum_{i=1}^W p_i =1 \Bigr) \,
\label{SBG}
\end{equation}
where $k$ is a conventional positive constant (in physics, typically taken to be the Boltzmann constant~$k_B$). This expression enables, as is well known, the construction of what is usually referred to as (BG) statistical mechanics, a theory that is notoriously consistent with thermodynamics. To be more precise, what is well established is that the BG thermostatistics is {sufficient} for satisfying the principles and structure of thermodynamics. Whether it is or not also {necessary} is a most important question that we shall address later on in the present paper. This crucial issue and its interconnections with the Boltzmann and the Einstein viewpoints have been emphatically addressed by \hl{E.G.D.} Cohen in his acceptance lecture of the 2004 Boltzmann Award \cite{Cohen2005}.

On various occasions, generalizations of the expression (\ref{SBG}) have been advanced and studied in the realm of information theory. In 1988, \cite{Tsallis1988} (see also \cite{CuradoTsallis1991,TsallisMendesPlastino1998}) the generalization of the BG statistical mechanics itself was proposed through the expression:
\begin{equation}
S_{q}=k \frac{1-\sum_{i=1}^W p_i^q}{q-1} = k \sum_{i=1}^W p_i \ln_q \frac{1}{p_i} \;\;\;\Bigl(\sum_{i=1}^W p_i =1; q \in \mathbb{R}; S_1=S_{BG} \Bigr) \,
\label{Sq}
\end{equation}
where the $q$-logarithmic function is defined through $\ln_q z \equiv \frac{z^{1-q}-1}{1-q}$ ($\ln_1 z =\ln z$). Its inverse function is defined as $e_q^z \equiv [1+(1-q)z]^{\frac{1}{1-q}} \;(e_1^z=e^z)$. Various predecessors of $S_q$, $q$-exponentials and~$q$-Gaussians abound in the literature within specific historical contexts (see, for instance, \cite{Tsallis2009} for a list with brief~comments). 

\section{Additive Entropy versus Extensive Entropy}
\unskip
\subsection{Definitions}
An entropic functional $S(\{p_i\})$ is said to be {additive} (we are adopting Oliver Penrose's definition~\cite{Penrose1970}) if, for any two {probabilistically independent} systems $A$ and $B$ (i.e., $p_{i,j}^{A+B}=p_i^A p_j^B, \; \forall (i,j)$),
\begin{equation}
S(A+B)=S(A) + S(B) \;\;\; [S(A+B) \equiv S(\{p_{i,j}^{A+B}\}); \, S(A) \equiv S(\{p_{i}^{A}\}); S(B) \equiv S(\{p_{j}^{B}\})] \,
\label{additive}
\end{equation}

It can be straightforwardly proven that $S_q$ satisfies:
\begin{equation}
\frac{S_q(A+B)}{k}= \frac{S_q(A)}{k} + \frac{S_q(B)}{k} + (1-q) \frac{S_q(A)}{k} \frac{S_q(B)}{k}\, 
\label{additiveq}
\end{equation}

Consequently, $S_{BG}=S_1$ is additive, whereas $S_q$ is non-additive for $q \ne 1$. 

The definition of extensivity is much more subtle and follows thermodynamics. A specific entropic functional $S(\{p_i\})$ {of a specific system} (or a specific class of systems, with its $N$ elements with their corresponding correlations) is said to be {extensive} if:
\begin{equation}
0< \lim_{N\to\infty}\frac{S(N)}{N} <\infty \,
\label{extensive}
\end{equation}
i.e., if $S(N)$ grows like $N$ for $N>>1$, where $N \propto L^d$, $d$ being the integer or fractal dimension of the system, and $L$ its linear size.

Let us emphasize that determining whether an entropic functional is additive is a very simple mathematical task (due to the hypothesis of independence), whereas determining if it is extensive for a~specific system can be a~very heavy one, sometimes even intractable.

\subsection{Probabilistic Illustrations}
If all nonzero-probability events of a system constituted by $N$ elements are equally probable, we have $p_i=1/W(N),\forall i$.

In that case, $S_{BG}(N)=k\ln W(N)$ and $S_{q}(N)=k\ln_q W(N)$.

Therefore, if the system satisfies $W(N) \propto \mu^N\,(\mu>1;\,N\to\infty)$ (e.g., for independent coins, we have $W(N)=2^N$), referred to as the {exponential class}, we have that the additive entropy $S_{BG}$ is also extensive. Indeed, $S_{BG}(N) \propto N$. For all other values of $q \ne 1$, we have that the non-additive entropy $S_q$ is nonextensive.

However, if we have instead a system such that $W(N) \propto N^\rho\,(\rho>0;\,N\to\infty)$, referred to as the {power-law class}, we have that the non-additive entropy $S_q$ is extensive for: 
\begin{equation}
q=1-\frac{1}{\rho} \;\;(\rho>0)\,
\end{equation} 

Indeed, $S_{1-1/\rho}(N) \propto N$. For all other values of $q$ (including $q=1$), we have that $S_q$ is nonextensive for this class; the extensive entropy corresponding to the limit $\rho \to\infty$ precisely is the additive $S_{BG}$. 

Let us now mention another, more subtle, case where the nonzero probabilities are {not} equal~\cite{TsallisGellMannSato2005}. We consider a triangle of $N \; (N=2,3,4,...)$ correlated binary random variables, say $n$ heads and $(N-n)$ tails $(n=0,1,2,...,N)$. The probabilities $p_{N,n}$ ($\sum_{n=0}^N p_{N,n}=1\;, \forall N$) are different from zero only within a strip of width $d$ (more precisely, for $n=0,1,2,...,d)$) and vanish everywhere else. This~specific probabilistic model is asymptotically scale-invariant (i.e., it satisfies the so-called Leibniz triangle rule for $N\to\infty$): see \cite{TsallisGellMannSato2005} for full details. For this strongly-correlated model, the non-additive entropy $S_q$ is extensive for a unique value of $q$, namely: \begin{equation}
q=1-\frac{1}{d} \;\; (d=1,2,3,...)\,
\end{equation} 

We see that the extensive entropy corresponding to the limit $d\to\infty$ precisely is the additive $S_{BG}$.

These examples transparently show the important difference between entropic additivity and entropic extensivity. What has historically occurred is that, during 140 years, most physicists have been focusing on systems that belong to the exponential class, typically either non-interacting systems (ideal gas, ideal paramagnet) or short-range-interacting ones (e.g., $d$-dimensional Ising, XY~and Heisenberg ferromagnets with first-neighbor interactions). Since for this class, but not so for many~others, the additive BG entropic functional is also extensive, a frequent confusion has emerged in the understanding of very many people and textbooks, which has led, and is unfortunately still leading, to somehow considering additive and extensive as synonyms, which is definitively false ({\highlight{this error is so easy to make, such that, by inadvertence,} the book \cite{GellMannTsallis2004} by Gell-Mann and myself was entitled {Nonextensive Entropy}, whereas it should have been entitled {Non-additive Entropy}; obviously, we definitively regret this misnomer}).

Further classes of systems do exist, for example the {stretched exponential} one, for which other entropic functionals (e.g., $S_{\delta}$ \cite{TsallisCirto2013}) are necessary in order to achieve extensivity. Indeed, no value of $q$ exists such that $S_q(N) \propto N$ for this class. In fact, a plethora of entropic functionals are now available in the information-theory literature (see, for instance, \cite{Renyi1961,Varma1966,SharmaMittal1975,SharmaTaneja1975,SharmaTaneja1977,AczelDaroczy1975,LandsbergVedral1998,Landsberg1999,Curado1999,AnteneodoPlastino1999,CuradoNobre2004,Zaripov2014,HanelThurner2011a,HanelThurner2011b}).

 \subsection{Physical Illustrations}

The entropic index $q$ is to be determined from first principles, namely from the time evolution (in phase space, Hilbert space and analogous) of the state of the full system. This typically is an~analytically hard task. Nevertheless, this task has been accomplished in some few cases. Let us briefly review some of them:
 
\begin{enumerate}[leftmargin=*,labelsep=5mm]
\item	The logistic map at its Feigenbaum point;
\item	The entropy of a subsystem of a $(1+1)$-dimensional system characterized by a central charge $c$ at its quantum critical point; 
\item	The entropy of a subsystem of a $(1+1)$-dimensional generalized isotropic Lipkin--Meshkov--Glick model at its quantum critical point.
\end{enumerate}

For the logistic map $x_{t+1}=1-ax_t^2\; (0<a<2; \,t=0,1,2,...; \, x_t \in [-1,1]$, we have that a value of $q$ exists, such that $S_q$ asymptotically increases {linearly with time}, where the value of $q$ is dictated by the Lyapunov exponent being positive or zero, which in turn depends on the value of the external parameter $a$. To be more precise, we assume the interval $[-1,1]$ of $x$ divided into $W$ tiny intervals (identified with $i=1,2,...,W$); we then place in one of those intervals many $M$ initial conditions (with $M>>W$); and finally, we iterate the map for each of these initial conditions. The number of points $M_i(t)$ that are located at the $i$-th interval satisfy $\sum_{i=1}^W M_i(t)=M \,,\forall t$. We define next the probabilities $p_i(t) \equiv M_i(t)/M$, which enable the evaluation of the entropy $S_q(t)/k =\frac{1-\sum_{i=1}^W [p_i(t)]^q}{q-1}$. It can be shown that a unique value of $q$ exists such that $K_q \equiv \lim_{t \to \infty} \lim_{W \to \infty} \lim_{M \to \infty} \frac{S_q(t)/k}{t}$ is {finite}. For any value of $q$ above this special one, the ratio $K_q$ vanishes, and for any value of $q$ below this special one, the ratio $K_q$ diverges. 

For all values of $a$ such that the Lyapunov exponent $\lambda_1$ is positive (i.e., in the presence of {strong} chaos, where the sensitivity to the initial conditions $\xi \equiv \lim_{\Delta x(0) \to 0}\frac{\Delta x(t)}{\Delta x(0)}$ increases exponentially with time, $\xi = e^{\lambda_1 \,t}$), we have that $q=1$, and the ratio precisely equals the Lyapunov exponent (i.e., $K_1=\lambda_1$; {Pesin-like identity}).

In contrast, at the edge of chaos, i.e., for the value of $a$ where successive bifurcations accumulate (sometimes referred to as the Feigenbaum point), i.e., $a=1.401155189092. . .$, we have that the Lyapunov exponent vanishes, and consistently \cite{LyraTsallis1998,BaldovinRobledo2002}, 
\begin{equation}
q=0.244487701341282066198...
\end{equation}
(in fact, \hl{1018} exact digits are numerically known nowadays \cite{Broadhurst1999}; see \cite{Tsallis2009} for full details). At such special values of $a$, we verify that $\xi=e_q^{\lambda_q \,t}$, where a $q$-generalized version of the Pesin-like identity has been rigorously established \cite{BaldovinRobledo2002}. The edge of chaos of logistic-like maps provides a remarkable connection of $q$-statistics with multifractals \cite{LyraTsallis1998}. This is particularly welcome because the postulate of the entropy $S_q$ in order to have a basis for generalizing BG statistics was inspired precisely by the structure of multifractals. The present status of our knowledge strongly suggests that a BG system typically ``lives'' in a smoothly-occupied phase-space, whereas the systems obeying $q$-statistics ``live'' in hierarchically-occupied phase-spaces. 

Let us now address the entropy of an $L$-sized block of an $N$-sized quantum system at its quantum critical point, belonging to the universality class, which is characterized by a central charge $c$ (e.g., the~universality classes of the short-range Ising and the short-range isotropic XY ferromagnets correspond respectively to $c=1/2$ and $c=1$). It has been shown \cite{CarusoTsallis2008} that $S_q$ is extensive for:
\begin{equation}
q=\frac{\sqrt{9+c^2} -3}{c} \,
\end{equation}

We verify that $c\to\infty$ yields $q=1$ (BG).

Finally, let us address the generalized isotropic Lipkin--Meshkov--Glick model \cite{CarrascoFinkelGonzalezLopezRodriguezTempesta2016}, characterized by $(m,k)$, where $m$ is the number of states of the model (e.g., if the system is constituted by $s$-sized spins, we have $m=2s$, $s=1/2,1,3/2,...$), and $k$ ($k=0, 1, 2,...$) is the number of vanishing magnon densities. The entropy $S_q$ is extensive for:
\begin{equation}
q=1-\frac{2}{m-k}=1-\frac{2}{2s-k} \;\;\;(m-k=2s-k \ge 3; \, q \ge 1/3) \,
\end{equation}

Notice that, in the limit $s \to\infty$, $q=1$ (BG).

Numerical results are available as well in the literature. For example, for a random antiferromagnet with $s$-sized spins, we have \cite{SaguiaSarandy2010}:
\begin{equation}
q \simeq 1-\frac{1.67}{\ln (2s+1)} \,
\end{equation} 
 
Before we proceed with analyzing thermodynamical aspects, let us stress that we have addressed here two different types of linearities, the thermodynamical one (i.e., $S_q(N) \propto N$) and the dynamical one (i.e., $S_q(t) \propto t$). Although the nature of these linearities is different and even the values of $q$, which guarantee them, may be different (although possibly related), there are reasons to expect both to be satisfied on similar grounds: this question was in fact (preliminarily) addressed in \cite{TsallisGellMannSato2005b} and elsewhere.

\subsection{Renyi Entropy versus $q$-Entropy}

Let us address here a question that frequently appears in the literature, generating some degree of confusion. We refer to the discussion of Renyi entropy {versus} $q$-entropy on thermodynamical and dynamical grounds. The Renyi entropy \cite{Renyi1961} is defined as:
\begin{equation}
S_q^R \equiv \frac{\ln \sum_{i=1}^W p_i^q}{1-q} \;\;\;\Bigl(\sum_{i=1}^W p_i =1; q \in \mathbb{R}; S_1^R=S_{BG} \Bigr) \,
\end{equation}
hence:
\begin{equation}
S_q^R = \frac{\ln[1+(1-q)S_q/k]}{1-q} \,
\end{equation}

It is straightforward to verify that $S_q^R(S_q)$ is a monotonic function of $S_q$, $\forall q$. Consequently, {under~the same constraints}, the extremization of $S_q^R$ yields precisely the {same distribution} as the extremization of $S_q$ (in total analogy with the trivial fact that maximizing, {under the same constraints}, $S_{BG}$ or say $[S_{BG}]^3$ yields one and the same BG exponential weight). This mathematical triviality is at the basis of sensible confusion in the minds of some members of the community. Thermodynamics and statistical mechanics is much more than a mere probability distribution, and the reader has surely never seen, and this for more than one good reason, constructing a successful theory such as thermodynamics by using say $[S_{BG}]^3$ instead of $S_{BG}$.

To make things more precise, let us list now several important differences between $S_q$ and $S_q^R$ (see, for instance, \cite{Tsallis2009} and the references therein). \\
\begin{enumerate}[leftmargin=21pt,labelsep=7pt]
\item [(i)] {Additivity:} If $A$ and $B$ are two arbitrary probabilistically-independent systems, $S_q^R$ is additive, $\forall q$, whereas $S_q$ satisfies the non-additive property in Equation (\ref{additiveq}). 

\item [(ii)] {Concavity:} $S_q(\{p_i\})$ is concave for all $q>0$, whereas $S_q^R(\{p_i\})$ is concave only for $0<q \le 1$. Both $S_q$ and $S_q^R$ are convex for $q<0$. These properties have consequences for characterizing the thermodynamic stability of the system.

\item [(iii)] {Lesche stability:} $S_q$ is Lesche-stable $\forall q>0$, whereas $S_q^R$ is Lesche-stable only for $q=1$. Lesche stability characterizes the experimental reproducibility of the entropy of a system.

\item [(iv)] {Pesin-like identity:} For many physically important low-dimensional conservative or dissipative nonlinear dynamical systems with zero Lyapunov exponent, it is verified that, in the $t \to\infty$ limit, $S_q(t) \propto t$ for a unique special value of $ q\ne 1$. This linearity property for $t>>1$ is lost for $S_q^R(t)$; indeed, for those systems, it can be easily verified that $S_q^R(t) \propto \ln t \,(\forall q)$. No dynamical systems are yet known for which $S_q^R(t)$ is linear for $q \ne 1$. This linearity enables, $\forall q$, a natural connection with the coefficient (Lyapunov exponent for the $q=1$ systems), which characterizes the dynamically meaningful sensitivity to the initial conditions. 

\item [(v)] {Thermodynamical extensivity:} For various $N$-sized quantum systems, it can be shown that a fixed value of $q \ne 1$ exists, such that, in the $N \to\infty$ limit, $S_q(N) \propto N$, thus satisfying the necessary thermodynamic extensivity for the entropy. For those systems, $S_q^R(N) \propto \ln N\,(\forall q)$, which violates thermodynamics. For this statement, we have of course assumed that a (physically meaningful) limit $q \ne 1$ exists in the $N \to \infty$ limit. Various papers exist in the literature that focus on situations such that a phenomenological index $q$ can be defined, which depends on $N$ (see, for instance, \cite{ParvanBiro2005,ParvanBiro2010} and the references therein), but they remain out of the present scope, since their $N\to\infty$ limit yields $q=1$.

\item [(vi)] {The likelihood function that satisfies Einstein's requirement of factorizability coincides with the function, which extremizes the entropic functional of the system (currently, the inverse function of the generalized logarithm, which characterizes that precise entropic functional:} For $q=1$ systems, the~{factorizable} likelihood function is well known to be ${\cal W} \propto e^{S_{BG}/k}$, the exponential function being the inverse of $S_{BG}/k= \ln W$ (for equal probabilities), and for appropriate constraints, it {maximizes} the entropy $S_{BG}$. For $q \ne 1$, we have \cite{TsallisHaubold2015} ${\cal W} \propto e_q^{S_{q}/k}$, where the $q$-exponential function precisely is the inverse of $S_q/k = \ln_q W$ (for equal probabilities), and for appropriate constraints, it {extremizes} the entropy $S_{q}$. In contrast with this property, the factorizable likelihood function for the Renyi entropy is $e^{S_q^R}$, where the exponential function is the inverse of $S_q^R=\ln W$ (for equal probabilities), but it {differs} from the $q$-exponential function, which is the one that extremizes $S_q^R$.
These properties plausibly have consequences for the large deviation theory of these systems (see the discussion about this theory below).
\end{enumerate}

\section{Why Must the Entropic Extensivity Be Preserved in All Circumstances?} 
 
Since we are ready to permit the entropic functional to be non-additive, should we not also allow for possible entropic nonextensivity? This question surely is a most interesting one, but to the best of our understanding, the answer is {no}. Indeed, there exist at least two important reasons for always demanding the physical (thermodynamical) entropy of a given system to be extensive. One of them is based on the Legendre transformations structure of thermodynamics; the other one is so suggested by the large deviations in some anomalous probabilistic models where the limiting distributions are~$q$-Gaussians.

 \subsection{Thermodynamics}
 
This argument has been developed in \cite{Tsallis2009} and more recently in \cite{TsallisCirto2013} (which we follow now). We~briefly review this argument here.~Let us first write a general Legendre transformation form of a~thermodynamical energy $G$ of a generic $d$-dimensional system ($d$ being an integer or fractal~dimension):
\begin{eqnarray}
G(V,T,p,\mu,H, \dots)&=&U(V,T,p,\mu,H,\dots) - TS(V,T,p,\mu,H,\dots) \\ 
 &&+pV -\mu N(V,T,p,\mu,H,\dots) - HM(V,T,p,\mu,H,\dots)-\cdots
\label{eq:thermodynamics}
\end{eqnarray}
where $T, p, \mu,H$ are the temperature, pressure, chemical potential and external magnetic field
and~$U,S,V,N,M$ are the internal energy, entropy, volume, number of particles and magnetization.
We~may identify three types of variables, namely: 
(i) those that are expected to always be extensive ($S,V, N,M,\ldots$), i.e., scaling
with $V \propto L^d$, where $L$ is a characteristic linear dimension of the system
(notice the presence of~$N$ itself within this class);
(ii) those that characterize the external conditions under which the system is placed ($T,p,\mu,H,\ldots$), scaling with $L^\theta$;
and (iii) those that represent energies ($G,U$), scaling with $L^\epsilon$.
Ordinary thermodynamical systems are those with~$\theta=0$ and $\epsilon=d$;
therefore,~both the energies and the generically extensive variables scale with~$L^d$, and
there is no difference between Type (i) and (iii) variables, all of them being extensive in this case.
There are, however, physical systems where~$\epsilon = \theta + d$ with~$\theta \neq 0$.
Let us divide Equation~\eqref{eq:thermodynamics} by $L^{\theta + d}$, namely,
\begin{equation}
\frac{G}{L^{\theta+d}} = \frac{U}{L^{\theta+d}} - \frac{T}{L^\theta}\, \frac{S}{L^d} + \frac{p}{L^\theta} \frac{V}{L^d} - \frac{\mu}{L^\theta} \,\frac{N}{L^d}
-\frac{H}{L^\theta}\,\frac{M}{L^{d}} - \cdots \,
\label{eq:thermodynamics2}
\end{equation}

If we consider now the thermodynamical $L\to\infty$ limit, we obtain:
\begin{equation}
\widetilde{g} = \widetilde{u} - \widetilde{T} s + \widetilde{p} v- \widetilde{\mu}\, n
-\widetilde{H} m - \cdots
\label{eq:thermodynamics3}
\end{equation}
where, using a compact notation, $(\widetilde{g}, \widetilde{u}) \equiv \lim_{L\to\infty} (G, U)/L^{\theta + d}$ represent the energies,
$(s, v, n, m) \equiv \lim_{L\to\infty} (S,V,N,M) /L^d$ represent the usual extensive variables
and $(\widetilde{T}, \widetilde{p}, \widetilde{\mu}, \widetilde{H}) \equiv \lim_{L\to\infty} (T, p, \mu, H)/L^{\theta}$
correspond to the usually intensive ones.
For a standard thermodynamical system (e.g., a real gas ruled by a Lennard--Jones short-ranged potential, a simple metal, etc.)
we~have $\theta =0$ (hence, $(\widetilde{T},\widetilde{p},\widetilde{\mu},\widetilde{H}) = (T,p,\mu,H)$, i.e., the usual intensive variables),
and $\epsilon = d$ (hence, $(\widetilde{g}, \widetilde{u}) = (g,u)$, i.e., the usual extensive variables);
this is of course the case found in the textbooks of thermodynamics.

The thermodynamic relations~\eqref{eq:thermodynamics} and \eqref{eq:thermodynamics2} put on an equal footing the entropy~$S$,
the volume~$V$ and the number of elements~$N$, and the extensivity of the latter two variables is guaranteed by definition.
In fact, a similar analysis can be performed using~$N$ instead of~$V$ since
$V\propto N$.

\begin{figure}[H]
\centering
 \resizebox{0.70\columnwidth}{!}{\includegraphics{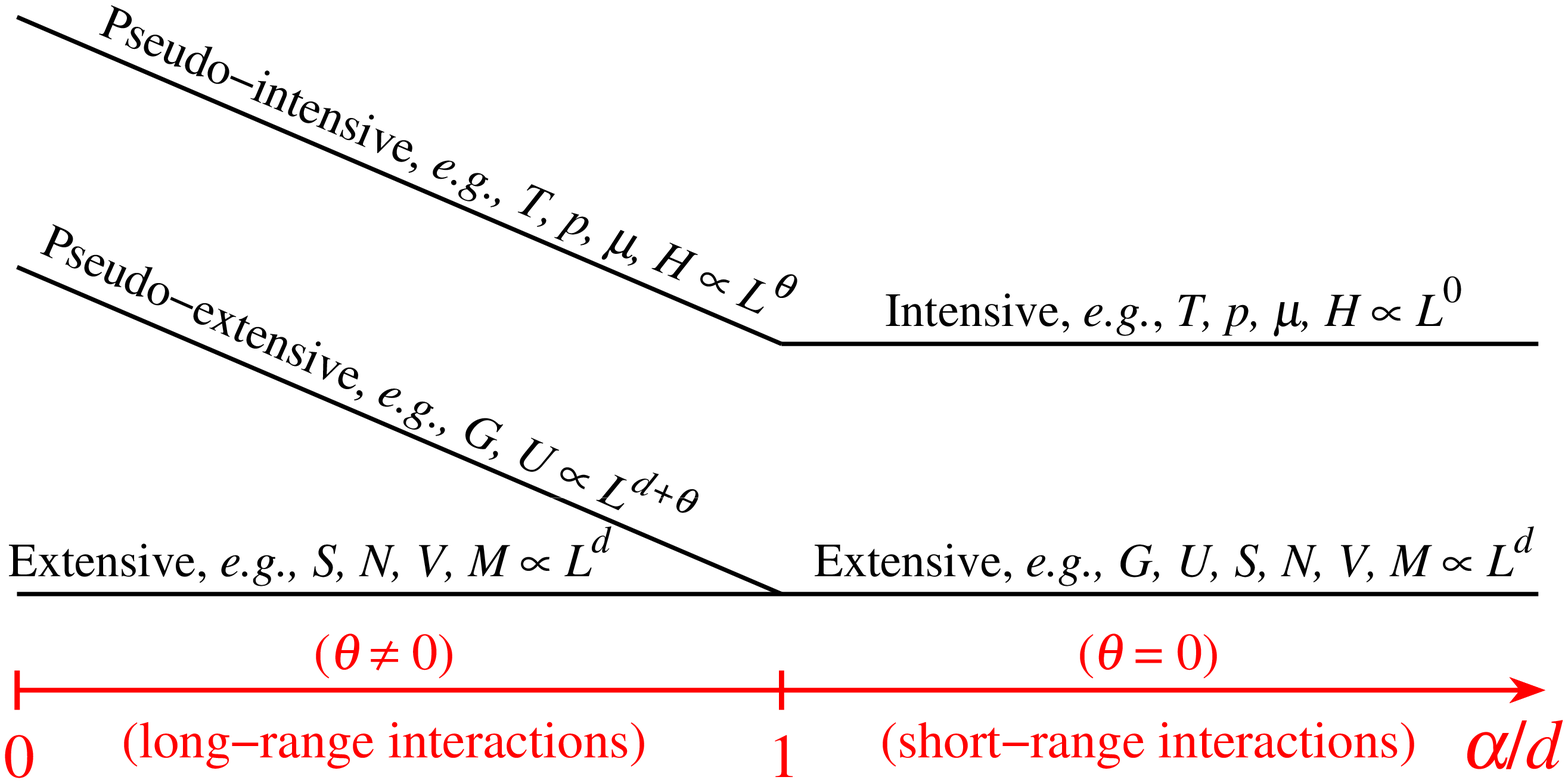} }
\caption{
Representation of the different scaling regimes of Equation~\eqref{eq:thermodynamics2} for classical $d$-dimensional~systems.
For attractive long-range interactions (i.e., $0 \leq \alpha/d \leq 1$, $\alpha$ characterizing the interaction range in a potential with the form $1/r^{\alpha}$),
we may distinguish {three} classes of thermodynamic variables, namely,
those scaling with $L^{\theta}$, named {pseudo-intensive} ($L$ is a characteristic linear length; $\theta$~is a system-dependent parameter),
those scaling with $L^{d+\theta}$, the {pseudo-extensive} ones (the energies),
and~those scaling with $L^{d}$ (which are always extensive).
For short-range interactions (i.e., $\alpha > d$),
we have $\theta=0$, and the energies recover their standard~$L^{d}$ extensive scaling,
falling in the same class of~$S$, $N$, $V$, etc., 
whereas~the previous pseudo-intensive variables become truly intensive ones (independent of $L$);
this is the region with {two} classes of variables that is covered by the traditional textbooks of thermodynamics. From \cite{TsallisCirto2013}.
}
\label{Fig:Pseudo}
\end{figure}

An example of a nonstandard system with~$\theta\neq 0$ is the classical Hamiltonian discussed in what~follows. 
We consider two-body interactions decaying with distance $r$ like $1/r^\alpha \;(\alpha \ge 0 )$.
For~this system, we have~$\theta =d-\alpha$ whenever $0 \le \alpha < d$ (see, for example, Figure~1 of~\cite{TamaritAnteneodo2000}).
This peculiar scaling occurs because the potential is not integrable,
i.e., the integral $\int_{\textrm{constant}}^\infty dr \,r^{d-1}\,r^{-\alpha}$
diverges for~$0 \le \alpha \le d$; 
therefore, the Boltzmann--Gibbs canonical partition function itself diverges.
Gibbs~was aware of this kind of problem and has pointed out~\cite{Gibbs1902} that whenever the partition function diverges, 
the BG theory cannot be used because, in his words, ``the law of distribution becomes~illusory''.
The~divergence of the total potential energy occurs for~$\alpha \le d$, which is referred to as long-range~interactions. 
If $\alpha > d$, which is the case of the $d=3$ Lennard--Jones potential, whose~attractive part corresponds to $\alpha=6$, 
the integral does not diverge, and we recover the standard
behavior of short-range-interacting systems with the~$\theta=0$ scaling.
Nevertheless, it is worth recalling that nonstandard thermodynamical behavior is not necessarily associated with long-range interactions
in the classical sense just discussed.
A meaningful description would then be long-range correlations (spatial or temporal),
because for strongly quantum-entangled systems, correlations are not necessarily connected with the interaction range.
However, the picture of long- versus short-range interactions in the classical sense, directly related to the distance~$r$,
has the advantage of illustrating clearly the thermodynamic relations~\eqref{eq:thermodynamics} and~\eqref{eq:thermodynamics2} for the different scaling regimes, as shown in Figure~\ref{Fig:Pseudo}.

To summarize this crucial subsection, we may insist that what is {thermodynamically relevant is that the entropy of a given system must be extensive, not that the entropic functional ought to be additive}. This~is consistent with the fact that Einstein's principle for the factorizability of the likelihood function is satisfied not only for the additive BG entropic functional, but also for nonadditive ones \cite{TsallisHaubold2015,SicuroTempesta2016}.
 
 \subsection{Large Deviation Theory} 
 
The so-called {large deviation theory} (LDT) \cite{Touchette2009} constitutes the mathematical counterpart of the {heart} of BG statistical mechanics, namely the famous canonical-ensemble BG factor $e^{-\beta {\cal H}(N)}~=~e^{-N [\beta h(N)]}$ with $h(N)\equiv {\cal H}(N)/N $. Since, for short-range interactions, $\beta h(N)$ is a thermodynamically-intensive quantity in the limit $N \to\infty$, we see that the BG weight represents an {exponential} decay with $N$. This~exponential dependence is to be associated \cite{Touchette2009,RuizTsallis2012,Touchette2013,RuizTsallis2013,RuizTsallis2015} with the LDT probability $P(N; x) \simeq e^{- N\,r_1(x)}$, where Subindex 1 in the {rate function} $r_1(x)$ will soon become clear. Since $r_1(x)$ is directly related to a relative entropy {per particle} (see, for instance, \cite{RuizTsallis2012}), the quantity $Nr_1(x)$ plays the role of \linebreak an {extensive}~entropy.

\begin{figure}[H]
\centering
\includegraphics[width=9.5cm]{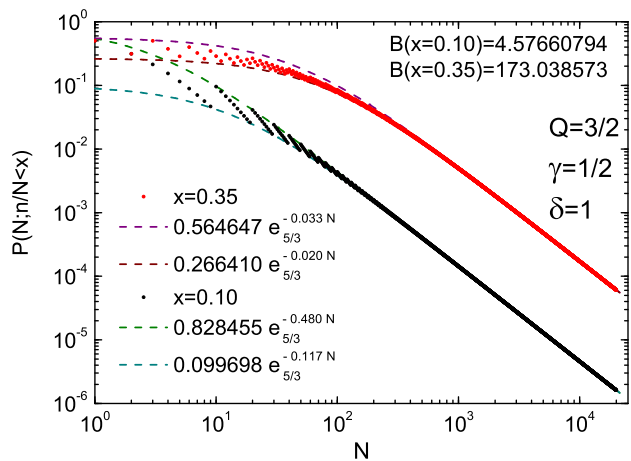}
\caption{Comparison of the numerical data (dots) of \cite{RuizTsallis2013} with $a(x)e_q^{-r_q N}$, where $(a(x),r_q(x))$ are positive quantities. From \cite{RuizTsallis2013}.
}
\label{comparison}
\end{figure}

\begin{figure}[H]
\centering
\includegraphics[width=8.5cm]{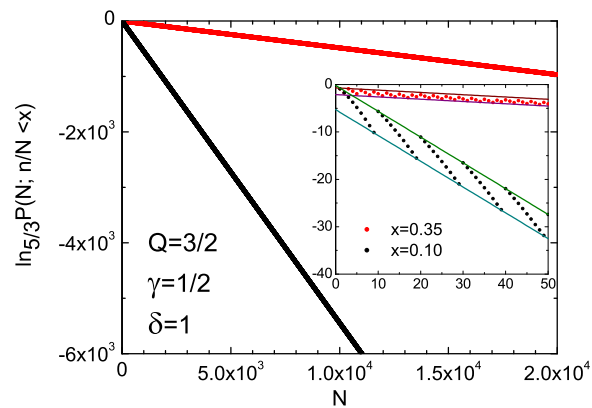}

\caption{The same data of Figure \ref{comparison} in ($q$-log)-linear representation. Let us stress that the {unique} asymptotically-power-law function, which provides straight lines {at all scales} of a ($q$-log)-linear representation, is the $q$-exponential function. The inset shows the results corresponding to $N$ up to 50. From \cite{RuizTsallis2013}.}
\label{comparisonqlog}
\end{figure}

If we focus now on, say, a $d$-dimensional classical system involving two-
body interactions whose potential asymptotically decays at long distance $r$
like $-A/r^{\alpha}$ $(A > 0; \alpha \ge 0)$, the canonical BG partition function converges
whenever the potential is integrable, i.e., for $\alpha/d > 1$ (short-range interactions), and diverges whenever it is non-integrable, i.e., for $0 \le \alpha/d \le 1$ (long-range interactions). The use of the BG weight becomes unjustified (``illusory'' in Gibbs words \cite{Gibbs1902} for, say, Newtonian gravitation, which in the present notation corresponds to $(\alpha,d)=(1,3)$; hence, $\alpha/d=1/3$)
in the later case because of the divergence of the BG partition function. We might therefore expect the emergence of some function
$f({\cal H}_N)$ different from the exponential one, in order to describe some specific stationary (or quasi-stationary) states differing from thermal equilibrium. The Hamiltonian ${\cal H}_N$ generically
scales like $N{\tilde N}$ with ${\tilde N} \equiv \frac{ N^{1-\alpha/d}-1}{1- \alpha/d} \equiv \ln_{\alpha/d} N$ (with the {$q$-logarithmic function} defined as $\ln_q z \equiv \frac{z^{1-q}-1}{1-q}; \,z > 0; \, \ln_1 z = \ln z$). Notice that ($N \to \infty$)
${\tilde N} \sim N^{1-\alpha/d}/(1-\alpha/d)$ for $1 \le \alpha/d < 1$,
${\tilde N} \sim \ln N$ for $\alpha/d = 1$
and ${\tilde N} \sim1/(\alpha/d -1)$ for $\alpha/d > 1$. The particular case $\alpha = 0$ yields ${\tilde N} \sim N$, thus recovering the usual prefactor of mean field theories. The quantity $\beta {\cal H}_N$ can be rewritten as $[(\beta {\tilde N}){\cal H_N} /(N{\tilde N})]N = [{\tilde \beta}{\cal H}_N /(N{\tilde N})]N$, where ${\tilde \beta} \equiv \beta {\tilde N} \equiv 1/k_B{\tilde T} = {\tilde N}/k_B T$ plays the role of an~intensive variable. The correctness of all of these scalings has been profusely verified in various kinds of thermal, diffusive and geometrical (percolation) systems (see \cite{RuizTsallis2013,Tsallis2009}). We see that, not only for the usual case of short-range interactions, but also for long-range ones, $[{\tilde \beta}{\cal H}_N/(N{\tilde N})]$ plays a~role analogous to an~intensive variable.
The {$q$-exponential function} $e_q^z \equiv [1 + (1 - q) z]^{\frac{1}{1-q}}$ ($e_1^z = e^z$) (and its associated $q$-Gaussian) has already emerged, in a considerable amount of nonextensive and similar systems, as the appropriate generalization of the exponential one (and its associated Gaussian). Therefore, it~appears as rather natural to conjecture that, in some sense that remains to be precisely defined, the LDT expression $e^{-r_1N}$ becomes generalized into something close to $e_q^{-r_q N}$ ($q \in {\cal R}$), where the generalized rate function $r_q$ is expected to be some generalized entropic quantity {per particle}. As shown in Figures \ref{comparison} and \ref{comparisonqlog} (see the details in \cite{RuizTsallis2013}), it is precisely this $e_q^{-r_q N}$ behavior that emerges in a strongly correlated nontrivial model \cite{RuizTsallis2012,RuizTsallis2013}. Since, as for the $q=1$ case, $r_q N$ appears to play the role of a total entropy, this specific illustration is consistent with an {extensive} entropy.

\unskip

\section{Further Applications and Final Words}

A regularly-updated bibliography on the present subject can be found at \cite{bibliography}. At the same site, a~selected list of theoretical, experimental, observational and computational papers can be found, as well. From these very many papers, let us briefly mention here a few recent ones. 

For those systems that may be well described by a specific class of nonlinear homogeneous $d=1$ Fokker--Planck equations, a prediction was advanced \cite{TsallisBukman1996} in 1966, namely the scaling $\mu= 2/(3-q)$, where $\mu$ is the exponent that characterizes the scaling between space and time (specifically the fact that $x^2$ scales like $t^\mu$) and $q$ is the index of the $q$-Gaussian, which describes the paradigmatic solution of the equation. Notice that $q=1$ yields the well-known Einstein 1905 result $\mu=1$ for Brownian motion. The prediction was experimentally verified (within a $2\%$ precision along an entire experimental decade), in 2015 \cite{CombeRichefeuStasiakAtman2015}, for confined granular material. It would be surely interesting to also verify, for higher-dimension confined granular matter, the $d$-dimensional generalization of that scaling, namely $\mu=\frac{2}{2 + d(1 - q)}$ \cite{MalacarneMendesPedronLenzi2001}; hence, once again $\mu =1$ for $q=1$.

For an area-preserving two-dimensional map, namely the standard map, it was neatly shown \cite{TirnakliBorges2016} how $q$-statistics, or BG statistics, or even a combination of both emerges as a function of the unique external parameter ($K$) of the map. This and various other emergencies of $q$-Gaussian and $q$-exponential distributions in many natural, artificial and social complex systems are most probably connected with $q$-generalizations of the central limit theorem (see, for instance, \cite{MoyanoTsallisGellMann2006,HilhorstSchehr2007,UmarovTsallisSteinberg2008,UmarovTsallisGellMannSteinberg2010,Hilhorst2010,JaureguiTsallis2011,JaureguiTsallisCurado2011,HahnJiangUmarov2010,JiangHahnUmarov2012,PlastinoRocca2012a,PlastinoRocca2012b,Budini2016}). 

Another $q$-statistical connection that certainly is interesting is the one with the so-called (asymptotically) scale-free networks. Indeed, their degree distribution has been shown in many cases to be given by $p(k) \propto e_q^{-k/\kappa}$ ($k$ being the number of links joining a given node), which plays the role of the Boltzmann--Gibbs factor for short-range-interacting Hamiltonian systems. This connection was already established in the literature since one decade ago (see, for instance, \cite{SoaresTsallisMarizSilva2005,ThurnerTsallis2005}). Moreover, it has been recently shown \cite{BritoSilvaTsallis2016} that neither $q$ nor $\kappa$ depend independently on the dimensionality $d$ and from the exponent $\alpha$ characterizing the range of the interaction, but, interestingly enough, only depend on the ratio $\alpha/d$. Very many papers focus on the degree distributions of (asymptotically) scale-free networks from a variety of standpoints. For example, an interesting exactly solvable master-equation approach is available in \cite{KullmannKertesz2001}. The novelty that we remind about in this mini-review is that the $q$-exponential degree distribution is here obtained from a simple entropic variational principle (under a constraint where the average degree plays the role of the internal energy in statistical mechanics).

High-energy physics has also been a field of many applications of $q$-statistics and related approaches, such as Beck--Cohen superstatistics \cite{BeckCohen2003} and Mathai's pathways (see \cite{Mathai2005,MathaiHaubold2007a,MathaiHaubold2007b,MathaiHaubold2008a,MathaiHaubold2008b} and the references~therein).~For example, a focus on the solar neutrino problem started long ago by Quarati and collaborators \cite{KaniadakisLavagnoQuarati1996,QuaratiCarboneGervinoKaniadakisLavagnoMiraldi1997,KaniadakisLavagnoQuarati1998,CoradduKaniadakisLavagnoLissiaMezzoraniQuarati1999} and has been revisited in several occasions, even recently \cite{MathaiHaubold2013,HauboldMathaiSaxena2014}. In~the area of particle high-energy collisions, an intensive activity is currently in progress.~It usually concerns experiments performed at LHC/CERN (ALICE, ATLAS, CMS and LHCb Collaborations) and RHIC/Brookhaven (STAR and PHENIX Collaborations). As typical illustrations of such measurements and their possible theoretical interpretations, let us mention \cite{BiroPurcselUrmossy2009,CleymansHamarLevaiWheaton2009,Cleymans2010,Biro2011,BiroVan2011,WongWilk2012,WongWilk2013,MarquesAndradeDeppman2013,BiroVanBarnafoldiUrmossy2014,WilkWlodarczyk2014,Deppman2014,WilkWlodarczyk2015a,RybczynskiWilkWlodarczyk2015,WilkWlodarczyk2015b,WongWilkCirtoTsallis2015a,WongWilkCirtoTsallis2015b,MarquesCleymansDeppman2015,DeppmanMarquesCleymans2015,Deppman2016}. A rich discussion about the thermodynamical admissibility of the possible constraints under which the entropic functional can be optimized is also present in the literature (see, for instance, \cite{TsallisMendesPlastino1998,FerriMartinezPlastino2005,ThistletonMarchNelsonTsallis2007,Tsallis2009,Abe2009,Biro2011}). 

Many other systems (e.g., related to those mentioned in \cite{CarideTsallisZanette1983,TirnakliTsallisLyra1999,TsallisLenzi2002,Tsallis2009b}) are awaiting for approaches along the above and similar lines. They would be very welcome. Even so, we may say that the present status of the theory described herein is at a reasonably satisfactory stage of physical and mathematical~understanding.

\vspace{6pt} 

\acknowledgments{I warmly dedicate this review to \hl{Arak M. Matha}i on his 80th anniversary. I am deeply indebted to Hans J. Haubold, whose insights and encouragements have made this overview possible, and also to the anonymous referees who, through highly constructive remarks, made possible an improved version of the present~manuscript. Finally, I acknowledge partial financial support from CNPq and Faperj (Brazilian agencies) and from the John Templeton Foundation (USA).}

\conflictofinterests{The author declares no conflict of interest.
}

\bibliographystyle{mdpi}

\renewcommand\bibname{References}

\end{document}